\documentclass[review,12pt]{elsarticle}

\usepackage{amssymb}

\journal{}

\usepackage{color}

\begin{document}

\begin{frontmatter}

\title{Econophysics: Bridges over a Turbulent Current}

\author[1]{Shu-Heng Chen}
\author[2]{Sai-Ping Li}
\address[1]{AI-ECON Research Center, Department of Economics, National Chengchi University,
Taipei, Taiwan. \\ chen.shuheng@gmail.com}
\address[2]{Institute of Physics, Academia Sinica, Taipei, Taiwan.  \\ spli@phys.sinica.edu.tw}

\begin{abstract}
In this editorial guide for the special issue on econophysics, we give a unique review of
this young but quickly growing discipline.  A suggestive taxonomy of the development is proposed by making a distinction
between classical econophysics and modern econophysics. For each of these two stages of
development, we identify the key economic issues whose formulations and/or treatments have
been affected by physics or physicists, which includes value, business fluctuations, economic
growth, economic and financial time series, the distribution of economic entities, interactions of
economic agents, and economic and social networks.  The recent advancements in these issues of modern
econophysics are demonstrated by nine articles selected from the papers presented at the {\em Econophysics
Colloquium 2010} held at Academia Sinica in Taipei.
\end{abstract}


\end{frontmatter}

% \linenumbers

\section{Introduction}

Despite their very different ages, physics and economics have been developed and extended
along the two sides of the same river for a long time.  Crossing the river signifies the efforts made to connect
the side of physics with the side of economics, or more generally, the side of the natural sciences
and the side of the social sciences.  More than one century ago, crossing the river had already started,
but over the years, particularly in recent years, the scale and organization of the crossings
have changed, from individuals to communities and from traveling to immigrating. To facilitate
such a massive crossing, bridges have also been built over the river.

The academic community currently known as {\em econophysics} can be regarded as an emerging
society after these crossings and the ensuring immigration.  All organized conferences and journals
(publications) related to this community are bridges.\footnote{Conferences regularly held on
econophysics include {\em Applications of Physics in Financial Analysis} (APFA), {\em
Econophysics Colloquium}, and {\em Econophys-Kolkata}. }  This special issue on econophysics
is one of these bridges and there are many bridges of this kind that have been built before us. Our
limited survey shows that there have already been eleven special issues published by journals
since the late 1990s. In chronological order, they are
\begin{itemize}
\item
{\em Physica A} 269(1) \citep{mantegna_1999},
\item
{\em International Journal of Theoretical and Applied Finance} 3(1) \citep{bouchaud_2000},
\item
{\em European Physical Journal B} 20(4) \citep{ausloos_2001},
\item
{\em European Physical Journal B} 27(2) \citep{richmond_ausloos_dacorogna_2002},
\item
{\em Physica A} 344(1) \citep{holyst_nowak_2004},
\item
{\em Physica A} 382 \citep{carbone_2007}
\item
{\em European Physical Journal B} 55(1) \citep{di_matteo_aste_2007},
\item
{\em Journal of Economic Dynamics and Control} 32(1) \citep{farmer_lux_2008},
\item
{\em Complexity} 14(3) \citep{shubik_smith_2009},
\item
{\em Science and Culture} 76(9-10) \citep{chakrabarti_chakraborti_2010}, and
\item
{\em AUCO Czech Economic Review} 4(3) \citep{vosvrda_2010}.
\end{itemize}
Several reviews of the development of econophysics have been nicely written by both
economists and physicists in the editorial guides of these special issues.
However, most of these reviews are not written in the journals to which economists usually
subscribe, and this special issue is one of the few exceptions.
Therefore, we feel inclined to start with a brief and unique review of the
background for a presumably very different group of readers.

\section{Economics and Physics: Their Interplay} \label{}

To begin with an interdisciplinary subject like econophysics, one naturally inquires
as to what parts of economics and what parts of physics are involved. If the fundamental pursuit is:
{\em whether we can understand economic phenomena by using the tools which we use to understand
physical phenomena, then we still have to answer what these tools and phenomena are}. However, both
economics and physics are more than a hundred years old. A lot can happen when we get that old,
which may make it difficult to provide a simple answer. Not only does a single
big event, such as the financial crisis, have effects on what econophysics should be, but
also the different ``dynasties'' in the long history of economics and physics can complicate
our answer.

In the history of orthodox economics, there is classical economics, neoclassical
economics, new classical economics, and Post-Keynesian economics, not to mention the
existence of many heterodox alternatives. Something equivalent exists in the history of physics,
which extends from classical mechanics, statistical physics, and quantum mechanics to relativity theory, etc. The long path
of each may characterize the interplay of the two over several different stages, which may
not be time consistent.  In this regard, \cite{rosser_2006} has well pointed out
that ``the much-derided standard models of economics largely came from physics.
(Ibid, p. 228)''  This time-inconsistency problem also exists in the relationship between
physics and mathematics. ``If the deterministic mechanical mode of physical argumentation
was to be replaced by an alternative physical theory, some established areas of mathematics
were no longer connected to a generally accepted physical model.
(\citep{weintraub_2002}, p. 10).'' Therefore, without a holistic picture of the historical
development, a person's perception of the relationship between economics and econophysics may be limited
and partial \citep{schinckus_2010}.

In this editorial guide, we hope to give a flavor of such a historical background not
just in economics and physics, but also in an increasingly growing collection of interdisciplinary
studies currently evolving among scientific communities.  Hence, our review will not just be
limited to modern econophysics but will start with classical econophysics. The main distinction
between classical econophysics and modern econophysics or anything in between lies in the
interdisciplinary context within which the crossing between the two happens. Most of the crossings in
classical econophysics do not involve other disciplines except, of course, mathematics, which
can be simply characterized as {\em link (point-to-point) crossings}.  However, crossings in
modern econophysics normally involve one or several other disciplines, in particular, the
advent of the complex-system community, and are better characterized as {\em network
crossings}.  As we shall see, our organization of the review, therefore, roughly corresponds to the division between
the era without the neologism ``econophysics'' and the era with it
\footnote{While econophysics as signifying the kinship between the fields of economics and physics has a long history,
the term ``econophysics'' was not seen until the 1990s.}, or to what
Bertrand Roehner termed {\em pre-econophysics} and {\em institutional econophysics}
\citep{roehner_2002}.

\section{Classical Econophysics \label{sec_classical}}

      In this section, we review what we consider to be the classical econophysics.  In this stage,
there are at least three fundamental economic phenomena being studied under the influence of
physics.  The three phenomena are {\em value}, {\em economic fluctuations}, and {\em economic growth}.  The
physics being applied to these phenomena
include rational mechanics, energetics and thermodynamics. Each of these areas involves
a number of economists consecutively working for quite a horizon. While their work had been
influential in economics at the time, their significance was either absorbed and hence replaced
by their successors or has become rather limited in recent years. It is in this
sense that we refer to these phenomena as classical econophysics.\footnote{Hence, this definition is different
from the one given in \citep{cottrell_2009}. By the same criterion, in this section we do not
include Louis Bachelier (1870-1946), who introduced what was later known as the Brownian motion to
the speculative price dynamics. While many econophysicists would like to mention his work as
the origin of econophysics, the influence of Bachelier's work \citep{bachelier_1990} to economics
was rather limited in his lifetime.}

\subsection{Energy and Value \label{sec_energy_value}}

The interplay between physics and economics and the social sciences already existed in the
19th century. In his book {\em Physics of Social Phenomena: An Essay
on Human Development} published in 1835, Adolphe Quetelet (1796-1874) had already attempted to
search for some statistical laws underlying certain social phenomena\footnote{These two
original volumes of Quetelet are in French and have never been translated into English.  For the
English title used above, we follow Bertrand Roehner \citep{roehner_2010}.}, and at that time he called
this study {\em social physics}, which came at a time almost 150 years earlier than when the field
``{\em sociophysics}'' was claimed to be formally founded by Serge Galam
\citep{galam_2004}.\footnote{Unfortunately, as we shall see in Section \ref{sec_interactions},
Serge Galam in \citep{galam_2004} made little reference to the early important work
done at the same time by Wolfgang Weidlich and Gunter Haag \citep{weidlich_1983}.} As we shall
see later in Section \ref{sec_distribution}, this search for the universal law or distribution
governing social phenomena has constantly been the main driving drive behind econophysics and
sociophysics, and hence connects classical with modern econophysics.

The influence of physics on economics can be traced all the way back to the late 18th or the
early 19th from classical economics to neo-classical economics. Philip Mirowski, a
historian of economic thoughts, asserted in his series of publications how the core concepts
of classical and neo-classical economics, such as labor and value, were developed in parallel
with the development of physics at that time, such as force, work, motion and
energy.\footnote{While Mirowsky's view nowadays has been widely cited by econophysicists, it
remains controversial among historians of economic thought. For example, see
\cite{hollander_1989}.}
\begin{quote}
Classical economists made reference to Newtonian analogy in non-essential contexts..., but
they could not reconcile the inverse square law, the calculus of fluxions and other
Newtonian techniques with their overall conception of social processes. The rise of
energetics in physical theory induced the invention of neoclassical economic theory, by
providing the metaphor, the mathematical techniques, and the new attitudes toward theory
construction. Neoclassical economic theory was appropriated wholesale from mid-nineteenth
century physics; utility was redefined so as to be identical with energy.
(\citep{mirowski_1984}, p.366)
\end{quote}
Rational mechanics and energetics, either metaphors or frameworks, have been used in the
writings of the major neoclassical economists, including William Jevons (1835-1882),
Leon Walras (1834-1910), Francis Edgeworth (1845-1926), Vilfredo Pareto (1848-1923) and
Irving Fisher (1867-1947).  As an illustration,
in his book {\em The Principles of Sciences} \citep{jevons_1905},
Jevons wrote:
\begin{quote}
Life seems to be nothing but a special form of energy which is manifested in heat and
electricity and mechanical force. The time may come, it almost seems, when the tender
mechanism of the brain will be traced out, and every thought reduced to the expenditure
of a determinate weight of nitrogen and phosphorous. No apparent limit exists to the
success of the scientific method in weighing and measuring, and reducing beneath the
sway of law, the phenomena of matter and mind...Must not the same inexorable reign of
law which is apparent in the motions of brute matter be extended to the human heart?
(Ibid, pp. 735-736.)
\end{quote}

Among the leading neo-classical economists, only Alfred Marshall had a reservation for the
physics or energetics metaphors and praised the biological metaphors highly. This can be
found in many places in his publications. For example,
\begin{quote}
In this vital respect all sciences of life are akin to one another, and are unlike physical
sciences. And therefore in the later stages of economics, when we are approaching nearly to
the conditions of life, biological analogies are to be preferred to the mechanical, other
things being equal. (\citep{marshall_1898}, ibid, pp.43)
\end{quote}

\subsection{Oscillations and Business Cycles \label{sec_cycles}}

      The second important development of physics in economics is the use of mechanical
design to demonstrate physical phenomena which can enhance or inspire our understanding of
economic phenomena.
In the 1930s, the exemplar of a simple machine used to understand business cycles
was the {\em pendulum}. Tinbergen (1903-1994), under the influence of James Clerk
Maxwell (1831-1879),
took harmonic oscillation - the mathematical representation of the pendulum - as a starting point
for analyzing the business cycle \citep{boumans_1992}.  Ragnar Frisch (1895-1973), in his debate
with Joseph Schumpeter (1883-1950) on business cycle theory, built a new mechanical analogy that considered an oscillating
pendulum whose movement was hampered by friction to take into account the irregular flow of
innovations \citep{boianovsky_trautwein_2007}. These innovations do not influence the period of movement, but
are necessary to keep the oscillations surviving.

      The most influential metaphor in the history of business cycle theory comes
from the {\em rocking horse}, a model initially proposed by Knut Wicksell (1851-1926) \citep{wicksell_1918}.  Frisch used this simple machine to
illustrate the distinction between impulse and propagation phenomena in cyclical
movements of damped systems \citep{frisch_1933}.  Frisch's rocking horse consists of three equations, that relate
macroeconomic variables, such as consumption, production and the money supply. Frisch imagined
the economy to be a rocking horse hit by a club.  The model then brought physical knowledge to bear
on the problem, through the equation which described a pendulum being dampened by friction. Frisch chose
values for parameters to replicate the real business cycle. This pioneering work shaped the
fundamental questions to be pursued in the next
half century's study on business cycles, namely, ``What are the sources and propagation
mechanisms for the boom/bust patterns of economic fluctuations in modern economies?
(\citep{quah_1995}, p. 1595)''

The further development of the mechanical analogies has led to the idea that a model of an
economy can be developed by identifying an analogy between a fluid flow and a monetary
flow.  In 1949 and 1950, A. William Phillips (1914-1975), a then sociology
undergraduate, and Walter Newlyn (1915-2002)
built a hydraulic-mechanical analogue  macroeconomic model, known as the {\em Phillips machine} or
{\em Moniac} \citep{phillips_1950}.\footnote{Phillips, however, is not the first
one to build an analogue computer for economic computation.  Irving Fisher had described
a hydraulic-mechanical analogue model for calculating the equilibrium price in 1891, and actually
built it in the 1920s, but it has been subsequently lost \citep{tobin_2005}.} It was an original 7 feet
$\times$ 5 feet $\times$ 3 feet representation of the macroeconomy. Oriented around monetary
stocks and flows represented by colored water flowing around plastic pipes, Moniac offered
the opportunity for policy simulation exercises \citep{lesson_2000}. An event to celebrate the 60th
Anniversary of the Phillips National Income Electro-Hydraulic Analogue Machine was held
by the {\em Algorithmic Social Science Research Unit} (ASSRU) at the University of Trento in December
2010.  Allan McRobie \citep{mcrobie_2011} has demonstrated a few more macroeconomic simulations using
the machine, and Kumararswamy Vela Velupillai \citep{velupillai_2011} has provided a deep reflection of the analogous
computing by recasting the Phillips machine in an era of digital computing.

\subsection{Thermodynamics and the Limits of Growth \label{sec_thermodynamics}}

     We shall close this section by walking from classical physics to thermodynamics and examine
its role in economics. Economics, since its very early stage, is a science of wealth creation.
A fundamental inquiry concerns the source of economic growth.  Whether there is a limit for economic
growth has long been a controversial issue in economics \citep{meadows_2004}.  In neoclassical economics, economic
growth is determined by technological progress, and as long as there is a constant influx of
new ideas, there is no a priori limit for growth. Even though natural resources have their
limits, technological advancements will constantly lead to new solutions, such as developing renewable
resources or the adoption of recycling or green technology.

     However, reservations to the above mainstream argument have existed for centuries.
The Physiocrats in the middle of the eighteenth century led by French economist
Francois Quesnay (1694-1774) argued that the economic process was subject to certain natural laws which operated
independently of human free will.  While the influence of the Physiocrats in economics quickly decayed
after the middle of the 18th century, Rudolf Clausius' (1822-1888)  work on
the second law of thermodynamics (the law of maximum entropy) in 1850 and the formal
presentation of entropy in 1865 provides
a new formulation of the Physiocrats. Nicolas Georgescu-Roegen (1906-1994) \citep{georgescu_roegen_1986} has
documented a historical review of this development, which eventually led to a biophysical
approach to economics, and has been referred to as {\em bioeconomics} by Nicolas Georgescu-Roegen
\citep{cleveland_1987,gowdy_mesner_2011}.\footnote{For other applications of thermodynamics to economics, the
interested reader is also referred to \citep{burley_foster_1994}.}

      The influence of thermodynamics on economics has a long history. {\em Entropy} (or energy)
and {\em the second law of thermodynamics} (the law of maximum entropy) have not only been
fundamentally considered to characterize economic processes, but have also technically
contributed to the formalism of econometrics.  In the 1950s, against the backdrop of the Shannon information theory, physicist
Edwin Jaynes (1922-1998) had already formulated the entropy maximization principle as the foundation of
statistical inference \citep{jaynes_1957}. This principle has since been extensively applied by
statisticians and econometricians in their modeling \citep{hill_1997}.

\section{Modern Econophysics \label{sec_modern}}

     Modern econophysics has been led by several pioneers.  Eugene Stanley and the Boston School
that he led kicked off the area by focusing on the subject which was rich in data, i.e., finance, or
more specifically, financial time series.  As time went on, new concentrations were also formed,
which not only helped shape econophysics but also extended it more generally to sociophysics.
In parallel to Section \ref{sec_classical}, the reviews that follow are organized into
four groups, each corresponding to one major economic phenomenon. These four are
(1) nonlinear dynamics, (2) distributions, (3) interactions and (4) networks.  These four, of
course, are not entirely mutually exclusive. Some econophysics or sociophysics applications
belong to more than just one of the four.

\subsection{Nonlinear Dynamics \label{sec_nonlinear dynamics}}

\subsubsection{Macroeconomic Dynamics}

      A long time before a large group of physicists had worked on the non-linearity of time
series or on
the non-linear economic dynamics, economicsts had already devoted themselves for decades to this
area in seeking to understand business cycles, financial markets and the instability of the capitalist
economy (also see Section \ref{sec_cycles}).  In macroeconomics, the literature on
non-linear business cycles, also known as endogenous business cycles, started in the middle
of the twentieth century with the help of economists, such as Nicholas Kaldor (1908-1986) \citep{kaldor_1940},
John Hicks (1904-1989) \citep{hicks_1950} and Richard Goodwin (1913-1996)
\citep{goodwin_1951}.  With the presence of the nonlinearity of certain
basic functional relationships within the system and lags in the feedback mechanism, these
non-linear models were able to demonstrate that aperiodic or periodic cycles are basically
inherent in the market economy, which can persist even without exogenous shocks.  These models,
however, fell out of flavor from the late 1950s onwards,
and the revival of the interests in them were not seen until the ``chaos wave''
came to economics in the early 1980s.

      From Henri Poincare (1854-1912) to Edward Lorenz (1917-2008), there are many different
intellectual origins of chaotic dynamics in its long history of development. Many of them arise
because of problems in physics, such as the three body problem, turbulence in fluid motion, and
nonperiodic oscillation in radio circuits.  Inspired by the study of deterministic
chaos and non-linear dynamics in mathematical physics and other disciplines, economists' interests
in non-linear economic models resurged. Since the early 1980s various aspects of non-linear mathematics have been
applied to theoretical and empirical economic models to study the macroeconomic phenomena
related to aperiodic cycles, strange attractors, bifurcation, phase transition,
multi-equilibria, path dependence and hysteresis effects.  A comprehensive collection of the early development
has been documented in \citep{benhabib_1992}.

\subsubsection{Non-Linear Time Series \label{sec_nonliear time series}}

In addition to macroeconomic dynamics, economic time series as the empirical counterpart of
dynamic economic theory have also been studied in depth in light of nonlinear dynamics,
with the ``chaos wave'' having accompanied a wave of the non-linear time series.  Therefore, the
interplay between economics and physics is not limited to macroeconomics, but also
econometrics, in particular, financial econometrics.

In the early 1990s, the Box-Jenkins paradigm (or the equivalent state-space
approach) and the vector auto-regression (VAR) models became well established in
textbooks on linear time series analysis, and new research directions for economic and
financial time series were nonlinear by nature.  Economists began to equip themselves with
various new techniques to tackle the non-linear properties in their data. New techniques
included non-linear (extended) Kalman filtering, threshold auto-regression, non-linear VAR,
chaotic dynamics, rescaled range analysis (Hurst exponents) and wavelets.  Some of these tools,
again, have physical origins; hence joint efforts between economists
and physicists were also observed in these works.\footnote{For the literature which documents
the advancement of non-linear economic time series and its possible influences by physics,
interested readers are referred to \cite{tong_1993,frank_stengos_1998,franses_dijk_2000,crowley_2007}.}

Accompanying or within this wave of non-linear time series is the increasing skepticism
regarding the Gaussian distribution, or what economists used to call the normal distribution.  In
fact, long before this wave, the fundamental work of Benoit Mandelbrot (1924-2010) and
Eugene Fama in the 1960s had already been strongly in favor of the stable Paretian distribution
as a model for the unconditional distribution of asset returns
\citep{mandelbrot_1963,fama_1963,fama_1965}.\footnote{However, the empirical studies
that test the stable distribution hypothesis in economics and finance continue to be a challenging
issue. See, for example, \cite{brock_1999}.} Inevitably, this skepticism also led to the
increasing reliance on non-linear models.\footnote{This is naturally so
because, by taking the conditional expectations as an example, it can be shown that its linear
form is no longer guaranteed if the multivariate Gaussian is violated.}  Empirical evidence
of financial returns not lending support to the Gaussian distribution have piled up since the 1980s;
as a consequence, in the 1990s the use of non-Gaussian distributions in financial time series
gradually became the rule rather than the exception \citep{mcculloch_1996,rachev_2003,jondeau_2006}, and some
pioneering work in econophysics has also been devoted to this direction, as we shall see in
Section \ref{sec_distribution}.

Equally important is the skepticism on the probabilistic independence of the asset return,
which is the backbone of the orthodox finance theory, namely, the efficient markets
hypothesis.\footnote{This can persistently be an issue under debate. Burton Malkiel, the
author of {\em A Random Walk Down Wall Street}, has a few excellent surveys
on this subject. He claims that stock market prices are far more efficient and far less
predictable than many academic papers would have us believe, and professional investment
managers, both in the U.S. and abroad, do not outperform their index benchmarks.
\citep{malkeil_2003, malkeil_2005}} In the 1980s, financial economists had already noticed that the
auto-correlation functions of several simple transformations such as the absolute value of
the return and the square of the return, also known as volatility, did not comply with the
independence assumption.  What has been particularly important at this stage is the development
of the nonlinear econometric test which can help distinguish the non-linear dependence from
linear independence. The most well-cited econometric test is the Brock-Dechert-Scheinkman
test or, simply, the BDS test \citep{bds_1996}.  This test is built upon a
correlation-dimension test developed by two physicists, Peter Grassberger and Itamar Procaccia,
and hence is also known as the Grassberger-Procaccia test \citep{gp_1983}.  Many financial
time series are found to be non-linear dependent through the BDS test.

      One fundamental work related to non-linear dependence is \citep{engle_1982}. Robert Engle
in 1982  proposed a model which demonstrates how the
volatility of returns is time-dependent and hence its future can be predicted from the past.
This celebrated model, known as ARCH (Autoregressive Conditional Heteroskedasticity), and its
generalizations, extensions and variations have quickly spread throughout financial
econometrics during the 1980s and 1990s.\footnote{For a survey on the
univariate ARCH-type models, see \citep{pagan_1996}, and for a survey on the multivariate
ARCH-type models, see \citep{bauwens_2006}.}  This new class of volatility models has
had a dramatic impact on option pricing. The conventional option pricing theory, the well-known
Black-Scholes model, is built upon the constant variance framework of the geometric Brownian motion.
Now, in light of the new empirical evidence that volatility
is not constant but time-dependent, addition work has been conducted to take this violation
into account. Recent advances in option pricing can be characterized as
the corrections of the biases associated with the Black-Scholes models with the presence of
different volatility assumptions.  This research issue was already initiated by mathematical
economists or econometricians \citep{duan_1995}, but later on it also attracted the interest of
econophysicists \citep{mccauley_2003,mccauley_2004}.

This paradigm shift characterized by the device of non-Gaussian distribution and non-linear
dependence in fact happened a little earlier before physicists began to examine the tail
behavior of all indices in light of the power law or scale-free distribution
\citep{mantegna_stanley_1995,levy_solomon_1998}.  Hence, when entering the 2000s, economists
and physicists developed a converging research interest in this regard.
As many earlier articles have pointed out, ``conflicts'' or ``prejudices'' always
exist between immigrants and local residents \citep{rosser_2006}, but, to the best of our
understanding, the area ``non-linear dynamics and non-linear time series'' is probably the
sub-community which enjoys the most intensive communication. As a result, the joint
efforts of economists and physicists have contributed to a long list of {\em stylized facts} in
financial time series, that cover the characteristics of returns, volatilities, trading volumes,
and trading breaks of both low-frequency and high-frequency data.\footnote{\cite{cont_2001}
initializes a list of the stylized facts of financial time series. \cite{chen_chang_du_2012}
continues adding a few more to make it more comprehensive.}

\subsection{Distribution \label{sec_distribution}}

The second theme of modern physics is the distribution behavior of economic activities.
As we have seen in Section \ref{sec_energy_value} when mentioning Adolphe Quetelet, the study of
the distribution of economic activities seems to provide the strongest motivation
for the search for universal methods for scientific inquiry. This has been
further elucidated by Herbert Simon (1916-2001), who tries to identify a class
of distributions which are applicable to rather extensive social and natural
phenomena \citep{simon_1955}. These distributions include two skewed distributions,
which econophysicists frequently cited, one being the Pareto distribution of income
and the other the Zipf distribution of the frequency of the occurrence of words.
Simon's pioneering work provides an empirical foundation for one kind of universality which
motivates physicists to work on economics or the social sciences.

The skewed distribution studied by Simon has been constantly followed and extended by
others in the economic literature and, recently, also pursued by the econophysics community.
The development of this literature can be roughly characterized by three directions.  First,
the skewed distributions are found to be applicable to many more economic variables. In
addition to income and wealth, they have also been applied to firm size, asset returns,
city size, film returns, innovation size, etc.\footnote{See \cite{gabaix_2008} and
\cite{rosser_2008} for a long list of these extensions.}
There are lot of breakthroughs during this period worth mentioning, but due to limitations of space,
we only mention three, namely, the work done by M. F. M. Osborne, Benoit Mandelbrot, and the Boston School led by
Eugene Stanley.

Osborne is considered to be the first to introduce the
lognormal stock pricing model \citep{osborne_1959} and independently apply the Brownian motion
to percentage changes in the stock price.  Physicist Joseph McCauley suggested that Osborne should be
honored as the first econophysicist \citep{mccauley_2006}. Mandelbrot, in his study on
the pattern of speculative prices (cotton in this case), first introduced
the term {\em Pareto-Levy distribution} or {\em stable distribution} to economics
\citep{mandelbrot_1960,mandelbrot_1963_b}. The Boston school first demonstrated the
applicability of the scaling law to financial indices \citep{mantegna_stanley_1995}.

The second direction concerns the statistical or econometric techniques chosen to identify the appropriate skewed
distribution among many possibilities. In addition to the frequently-cited Pareto and Zipf distributions,
there are lognormal and Yule distributions plus many generalizations of them that are often
considered. These distributions may look similar by simply eye-browsing.  Therefore, the distinction among them requires
deliberate statistical analysis.  A concern for the insufficiency of the technical
rigorousness has been recently brought up in \cite{gallegati_2006}, which triggers another
intensive communication between economists and physicists.\footnote{\cite{gallegati_2006} can be
read as criticisms of the modern econophysics contributed by physicists.  Four criticisms have been
outlined that are not just limited to the empirical work of the power law, but that include several others. This
article is so ``inspiring'' that it has received tremendous feedback from physicists. See,
for example, \citep{mccauley_2006,di_matteo_aste_2007}.}

One important reason for distinguishing different skewed distributions is that they may be
associated with different underlying mechanisms. An example shown by Simon is that depending
on whether the birth process is involved, one can have either a Yule distribution or lognormal
distribution \citep{simon_bonini_1958}.  Therefore, the third development in this line is to
build the theory or offer explanations that underlie these distributions.  The mechanism
proposed by Simon is a cumulative advantage mechanism, which is based on an early work by a
British statistician Udny Yule (1871-1951).  Later on, this mechanism, also known as
preferential attachment, had a great influence on the literature of the physics of complex
networks (Section \ref{sec_networks}).\footnote{Other recent reviews of these mechanisms
can be found in \citep{gabaix_2008,mitzenmacher_2008}. }
Since what we are dealing with involves the evolution of the distribution of economic
activities (income or firm size) over time, a general mathematical framework for describing this
evolution is the familiar {\em master equation} which originated from {\em statistical
physics}. A related alternative to statistical physics is {\em agent-based modeling}.
These two approaches are considered highly complementary in current econophysics in dealing with economic and
social interaction, the subject to which we now turn.

%Coming to the era of modern econophysics, the phenomena studied by Pareto, Zipf, Simon and
%Manderbrot are known as power laws or scale-free distributions and its relevancy to
%financial markets have been intensively founded by Eugen Stanley and many others. The class of
%scale-free distribution can be shown to satisfy entropy maximization giving some boundary
%conditions.  Therefore, the consistency with this entropy maximization principle provides a
%natural bridge between phenomena in economics and those in thermodynamics and mean field theory.

\subsection{Social Interactions \label{sec_interactions}}

Economics, in its mainstream, has for quite a long time been studied with the device of one
single agent, normally known as the {\em representative agent}.  This abstraction of the
macroeconomy or the market economy, as a highly decentralized system composed of interacting
heterogeneous agents, has been considered to be rather unsatisfactory for different schools of
economists in recent years \citep{kirman_1992,hartley_1997}. The {\em aggregation problem}
characterized as the summation over a set of interacting heterogeneous agents has been simply
assumed away in these representative-agent macroeconomic models \citep{blundell_stoker_2005,
chen_huang_2010}.  Alternative macroeconomic models built upon heterogeneous agents or interacting
heterogeneous agents have been proposed \citep{delli_gatti_2008}.  They are generally known as
{\em agent-based computational economics}. It is based on this development that we
see the relevance of {\em statistical physics} to economics.

Statistical physics, originally developed from statistical thermodynamics, gives
us a picture of how microscopic particles act in the aggregate to form the macroscopic
world, given the forces between microscopic particles.  This basic pursuit for the understanding
of the relationship between micro and macro is in line with agent-based computational economics;
therefore, their interplay is a matter of time and degree.  In fact, econophysics, for many
physicists and economists, is simply just the application of statistical physics,
and not other branches of post-Newtonian physics, to economics \citep{farmer_lux_2008}.

The history of the application of statistical physics to economics can be traced back to an renowned Italian
physicist Ettore Majorana (1906-1938, missing). Thanks to the English translations provided by Rosario
Mantegna, one of Majorana's articles ``The value of statistical laws in physics and social
sciences'' has become available in the journal {\em Quantitative Finance} \citep{majorana_2005}.
Of course, the application of statistical physics to economics dates back to much earlier than this
rediscovery.  Hans Follmer is the pioneer in this direction. Follmer \cite{follmer_1974}
is the first to explicitly use an {\em Ising model} to model the social interactions of
consumers and the resultant random but {\em interdependent} preferences. He showed
that with the presence of even short range interaction the microeconomic characteristics may no
longer determine the macroeconomic phase.  Other pioneers include Wolfgang Weidlich,
Gunter Haag, and Masanao Aoki.

      Weidlich and Haag \citep{weidlich_1983} are probably the first to introduce the
use of the {\em master equation} to study social systems.  They built various social dynamic
models upon the master equation to describe several social behaviors such as opinion formation, migration, and
the settlement structure.
In this vein, Masanao Aoki continued to advocate the relevance of the statistical mechanism to
macroeconomic modeling.  In particular, he demonstrated how a number of macroeconomic and
industrial dynamic problems can be represented by the jump Markov processes and can be solved
with the use of a master equation (the {\em Chapman-Kolmogorov Equation}) and {\em Fokker-Planck
equations} \citep{aoki_1996}.  He went further to use this framework to establish a
microeconomic foundation for Keynes's principle of effective demand, and argued that the
long-run economic growth can be demand-driven rather than just supply-driven as held by the
conventional view \citep{aoki_yoshikawa_2006}.  Other pioneers include Steven Durlauf, William Brock
and Laurance Blume, who popularized the use of the {\em Gibbs-Boltzmann distribution} in economic models, or, more specifically,
in their proposed interaction-based discrete choice models \citep{durlauf_1999}.

      Other physical models applied to modeling social interactions in economics
include {\em cellular automata}, {\em kinetic models}, {\em percolation models}, and {\em minority
games}.

\paragraph{Cellular Automata}
Cellular automata were invented by John von Neumann (1903-1956) based on the design of
self-reproducing  automata. Von Neumann drew some of his inspiration from his colleague in the Manhattan
project, Stanislaw Ulam (1909-1984). This model has subsequently been extensively applied to
simulating social interactions. Ulam was studying the {\em growth of crystals} using a simple lattice network approach.
He suggested to von Neunmann as early as 1950 that simple cellular automata could be found in sets of local rules that
generated mathematical patterns in 2-D and 3-D space where global order could be reproduced
from local actions.  Cellular automata models were then used in economics to study pricing in
a spatial setting \citep{keenan_obrien_1993}, sentiment dynamics \citep{chen_1997} and
technological innovation \citep{leydesdorff_2002}, etc.

\paragraph{Kinetic Model}
The kinetic theory of gases was used in the study of wealth and income distribution.
In this model, money-exchange trading was treated like the elastic scattering process in physics.
This kinetic model of income distribution was first studied by John Angle during the
1980s, and was referred to differently as the {\em inequality process}.  Angle's inequality
process is motivated by the surplus theory of social stratification in economic anthropology,
rather than by anything in physics \citep{angle_1986}.  Later on in the 2000s, this model was independently studied
again by physicists Adrian Dragulescu and Victor Yakovenko, who cause the model to become well-known
among econophysicists \citep{dragulescu_2000,chatterjee_2000}.  In a series of studies, Arnab Chatterjee and Bikas Chakrabarti showed
how the wealth distribution can change from the Gibbs distribution to the Gamma distribution and
further to the Pareto distribution by manipulating different saving behavior \citep{chatterjee_2004,chatterjee_2005}.
The kinetic model,
therefore, becomes the most parsimonious model which is able to account for the empirical
phenomena of wealth distribution.  Some economists, however, are very critical of this
model partially due to its lack of a realistic description of economic
behavior \citep{gallegati_2006,yakovenko_2009}.

\paragraph{Percolation Models}
The percolation theory was invented by Paul Flory (1910-1985), who published the first percolation theory in 1941, to explain polymer
gelation \citep{flory_1941}.  The percolation theory has been applied by Rama Cont and Jean-Philipe Bouchaud to
study the herding effect in financial markets \cite{cont_bouchaud_2000}.  Their model known as the Cont-Bouchaud
model is probably the first agent-based model of a financial market built by
explicitly taking into account the network effect.\footnote{For a survey on agent-based models of financial markets, the interested
reader is referred to \citep{samanidou_2007}.}  Despite its physical origin, the
operation of this model can be interpreted mathematically as a {\em random graph} with a given
probability that determines the existence of a link between any two points of the graph.
This probability parameter, also called the percolation parameter, plays a critical role in
this model as determining the distribution of the cluster size and the fluctuation of the price. Many further
variations of this model and its application to other fields, such as marketing, have been
well surveyed in \citep{samanidou_2007}.

\paragraph{Ising Models}
Earlier we mentioned that Ising models had first been used by Follmer in economics. While
Ising models, cellular automata and percolation models originated from different physical
observations, an equivalence relationship among the three can be established
\citep{domany_kinzel_1984}.  After brief reviews of the applications of cellular
automata and percolation models, we shall do the same here for Ising models.  The Ising model
originated from the dissertation of Ernst Ising (1900-1998).  Ising studied a linear chain of
magnetic moments, which are only able to take two positions or states, either up or down, and which are
coupled by interactions between nearest neighbors.  This model is widely used, not just in
physics, but also in biology and the social sciences.  In economics, it has been used to model
financial markets \citep{iori_1999,iori_2002,sornett_zhou_2006} and tax
evasion \citep{zaklan_2009}.

\paragraph{Minority Games}

      The minority game is considered to be one of the most successful econophysics models, even from the economists' viewpoint
\citep{gallegati_2006}.  There are a few games which are very simple and parsimonious, yet they often help us to gain deep
insights from the study of them.  These games are not only strongly favored by game theorists, but also social scientists
in general. Several famous ones include the prisoner's dilemma game, the ultimatum game, and the outguessing game (also known as Keynes's
beauty contest).  Using the metaphor from Robert Axelrod, we can call them the {\em E coli} of the social sciences.  The minority game
is another such example, which is better known to physicists than economists.

      The game was first introduced in 1994 by Brian Arthur \citep{arthur_1994} and is known as the {\em El Farol Bar Problem}.
Without pricing signals and other central intervention in the use of the space in the El Farol Bar, Arthur asked whether customers
can self-coordinate the attendance rate such that the bar will be well, but not over, used. While this problem is in general related to
the provision and the use of public goods, Arthur's main concern had to do with the kind of social or market order that may have come out
of the bounded rationality of customers. For example, would and how often would the bar be overcrowded? Very quickly one can see the minority
position in this issue as referring to those who did not attend the overcrowded bar or those who attended a rather spacious bar. In 1997, two
physicists Damien
Challet and Yi-Cheng Zhang took the essential idea of the minority position and formalized the minority game \citep{challet_zhang_1997}.
The main interest in studying the
minority game was directed toward financial markets where the minority position may play a crucial role. While it is
still not entirely clear how successfully one can build an economically relevant financial market model using a minority game,
the minority game has been seen as a prototype for demonstrating the applications of statistical mechanics to
interacting agents \citep{coolen_2004,challet_2005}.

%For convenience, we shall call it the Majarona's proposition. The Majarona's proposition
%is that we don't know many details and neither are they not important.
%This thinking facilitates the wide acceptance of the statistical physical approach to social
%sciences, and has impacted upon the latter development of agent-based modeling in two directions,
%one at a mesoscopic level  and one at a microscopic level.
%Taking financial markets as examples, the former has led to the celebrated class of the
%H-type agent-based modeling of financial markets, and the latter has led to the extensive use
%of the entropy-maximization agents in artificial stock markets.

\subsection{Complex Networks \label{sec_networks}}

       Various interaction models which we have reviewed above, from cellular automata to Ising models,  are all
special kinds of networks in which physical distance plays an important role in determining the interactions among components.
However, there is a large class of networks in which the physical distance is either negligible or is not the only
important determinant.  The social network is a good example.  Long before it caught the eyes of physicists,
the social network had already drawn the attention of sociologists. In fact, the term social network was
first coined by John Barnes in 1954 \citep{barnes_1954}.  In the late 1960s, Stanley Milgram and his student Jeffrey Travers conducted
their famous small-world experiment and verified the six degrees of separation \citep{travers_1969}. In the early 1970s,
Mark Granovetter, the founder of modern economic sociology,
proposed a network property referred to as {\em weak ties} and showed its significance in the operation of job markets \citep{granovetter_1973}.
In the middle of the 1980s, various economic decisions based on network externalities, such as consumption externalities and the adoption of
technology, were studied by economists \citep{david_1985,katz_shapiro_1985}.

       However, it was only in the middle of the 1990s that economists began to provide a formal treatment of networks.  The seminal work
by Matthew Jackson and Asher Wolinsky \citep{jackson_1996} and Venkatesh Bala and Sanjeev Goyal \citep{bala_goyal_2000} pioneered a
game-theoretic approach to study the formation of social and economic networks.  This is about the same time that physicists, such as Duncan
Watts, Steven Strogatz, Albert-Laszlo Barabasi, and Reka Albert started to search for the organizational principle of complex networks and proposed
their {\em small-world network} and {\em scale-free network}, respectively \citep{watts_strogatz_1998,albert_barabasi_2002}. While these two
approaches are complementary, the econophysicists' approach is more data-driven and has uncovered the network structure of many large-scale
economic datasets. The contribution of econophysicists to economic and social networks can be roughly divided into three related dimensions: first,
the empirical construction of the economic networks; second, the analytical techniques underlying the constructions; and third, the pattern
discoveries of networks (statistical properties of networks).

      The idea of providing a network representation of the whole economy started with Quesnay's {\em Tableau Economique} in 1758 (see also Section
\ref{sec_thermodynamics}), which depicted the circular flow of funds in an economy as a network.  Quesnay's work later on inspired
the celebrated {\em input-output analysis} founded by Wassily Leontief (1905-1999) in the 1950s \citep{leontief_1951}, which was further generalized
into the {\em social accounting matrices} by Richard Stone (1913-1991) \citep{stone_1961} in the 1960s. This series of development forms the backbone of
{\em computable general equilibrium analysis}, a kind of applied micro-founded macroeconomic model, pioneered by Herbert Scarf in the
early 1970s \citep{scarf_1973}.  These earlier ``network representations'' of economic activities enable us to see the interconnections and interdependence of
various economic participants.  This ``visualization'' helps us to address the fundamental issue in macroeconomics, i.e., how disruption propagates
itself from one participant (sector) to others through the network.

     The era of globalization provides us with a new drive to study and explore economic networks in a global context, which is important for
addressing the timely issue of financial security and stability. Hence, tremendous efforts have been made over the last decade to construct
networks of various flows within the global economy, which offers alternative approaches to the network representation of the real economy.  These
networks range from the flow of commodities (exports and imports, {\em world trade web}) \citep{serrano_boguna_2003} to the flow of capital
(direct and indirect foreign investment, {\em investment networks} or {\em financial networks})
\citep{battiston_2007,hocnberg_2007,kogut_2007,song_2009}. An international economic
network can also be built upon the correlations of macroeconomic fluctuations using the techniques introduced below ({\em GDP network})
\citep{ausloos_2007}. In addition to the macroeconomic networks, various industrial networks have also been established.  These include the networks of
companies, firms and banks \citep{uzzi_1996,souma_2007,aoyama_2010}.

     To construct the networks above, some new techniques have been introduced by physicists, for example, the use of {\em minimum spanning trees}
by Rosario Mantegna \citep{mantegna_1999b} and the {\em thresholding approach} by Jukka-Pekka Onnela \citep{onnela_2004}.  These techniques allow
us to provide a network representation of correlation matrices, known as {\em correlation networks}.  When applied to financial data, these networks
provide investors with a new way of examining financial information or making investment decisions.  The correlation networks have been applied to examine
networks of different assets, such as equities \citep{mantegna_1999b,onnela_2004} and currencies \citep{mizuno_2006}.  Additional techniques
have been introduced to build cross-correlation networks; in this way, the network is associated with a law of motion and is endowed with
a dynamic interpretation \citep{aste_2006,ausloos_2007}.  The correlation networks can be considered to be an approach to a more
general attempt, i.e., to map time series data into networks.  There are other approaches being developed for this more general attempt, such as the
{\em visibility graph} \citep{lacasa_2008}.  Some features of time series, such as periodicity and fractal, can then be inherited and manifested through
different network topologies, such as regular networks and scale-free networks.

     The other important development is the more flexible and rich representation of networks.  The conventional binary network has been
extended to the weighted network, such as the correlation networks.  In addition, the single graph has been expanded to multigraphs \citep{souma_2007},
i.e., there can be multiple links between nodes. The heterogeneity of nodes is also taken into account and the characteristics of nodes are then
incorporated as part of the network construction through hidden variable mechanisms \citep{caldarelli_2002,garlaschelli_2004}.\footnote{See
also the related discussions in \citep{chen_2006}.}

      Finally, various properties of economic networks have been identified, including the small-world and scale-free characterization of economic
networks \citep{aoyama_2010}, the scaling laws \citep{battiston_2007,duan_2007}, giant components \citep{kogut_2007}, clustered structures
\citep{mizuno_2006}, and weak and strong ties \citep{fagiolo_2010}, etc. These findings may have far-reaching implications for survivability
\citep{uzzi_1996,hocnberg_2007}, security \citep{nagurney_qiang_2009}, efficiency and many other issues. However, the causes and consequences of
various network topologies in general remain a challenge.

\section{Article Synopsis}

Articles published in this special issue are selected from the papers presented in Econophysics Colloquium 2010. Econophysics
Colloquium has been a major series for econophysics.  It was started in Canberra (2005), then followed in Tokyo (2006),
Ancona (2007), Kiel (2008), and Erice (2009).   This one in Taipei had a total of 69 presentations, and 23 of them were submitted for
publication consideration in this special issue.  All papers were sent to two anonymous referees for two sets of reviews (the
original one and the revised one). Nine papers were finally accepted for inclusion in this issue.  They represent the advancements made in
various areas as reviewed in Section \ref{sec_modern}.

Among the nine papers, four document the continuing research on the direction as
summarized in Section \ref{sec_nonlinear dynamics}, particularly, in Section
\ref{sec_nonliear time series}.  Together they are contributions to return volatility,
the non-stationarity of financial time series, and portfolio strategies.
The article ``Quantifying Volatility Clustering in Financial Time Series''
by Jie-Jun Tseng and Sai-Ping Li proposes a novel measure of volatility clustering based on
a crucial but less well noticed pattern in financial time series, namely, the bumps
appearing in the non-linear autocorrelation function of returns.  The article ``Properties of Range-Based Volatility Estimators''
authored by Peter Monlar studies the statistics of the range-based estimator of volatilities and proposes a modified version
by taking into account the open jumps. An interesting finding is that returns normalized by their standard deviations, obtained
from the proposed range-based estimated volatility, are not fat-tailed but are approximately Gaussian.
Using high-frequency data, Takaaki Ohnishi, in his article ``On the Nonstationarity of the Exchange Rate Process'',
presents the evidence that the exchange rate is not strictly stationary.  He further found that the waiting time for the regime change
follows an exponential distribution.  The nonstationarity issue of the mean-variance of stock returns is also studied in the
paper ``Mixed Time Scale Strategy in Portfolio Management'' authored by Wenjin Chen and Kwok Yip Szeto.  There they construct
a portfolio based on both a long-term trend guided by financial principles and a short-term trend governed by the specific trading
mechanisms used.  This mixed time-scale portfolio is shown to have superior performance to the respective market index.

Two contributions are pertinent to Section \ref{sec_distribution}. The article ``Market Fraction Hypothesis: A Proposed Test'',
by Michael Kampouridis, Shu-Heng Chen and Edward Tsang, examines the distribution of strategies adopted by traders over time.
The fundamental question to pursue is whether the long-term distribution is uniform over the strategy space so that all strategies are equally
attractive or unattractive to traders. This behavior, coined as the market fraction hypothesis, can be regarded as an application of the entropy
maximization principle to market microstructure. Using empirical data from ten different financial markets, they are able to
characterize some features of short-term dynamics and long-term distributions related to the market fraction hypothesis. One of their findings is
that the extent to which the market fraction hypothesis is sustained depends on how coarse or fine is our differentiation of different
trading behavior.  In the paper entitled ``Patterns of Regional Travel Behavior: An Analysis of Japanese Hotel Reservation Data'',
by Aki-Hiro Sato,  a finite mixture of Poisson distributions is applied to study the tendency of regional travel behavior.  Data associated with four
tourist attraction areas in Japan are used to estimated the model.  The demand for and supply of
hotel rooms are characterized by means of the relationship between the average room prices and the probability of room availability.

There are three contributions related to Section \ref{sec_interactions}.  Two are devoted to agent-based financial markets, and one
is devoted to the kinetic model of wealth distribution.  Based on the taxonomy given in \cite{chen_chang_du_2012}, roughly speaking, there are two
types of agent-based financial markets, namely, the {\em H-type ones} and the {\em SFI (Santa Fe Institute) ones}. Both types are initiated and
developed by economists, and not physicists.  The agent-based financial model studied by Lukas Vacha, Jozef Barunik and Miloslav Vosvrda
in their article ``How Do Skilled Traders Change the Structure of the Market'', is an example of an H-type financial market.  Within
the Brock-Hommes framework \citep{brock_hommes_1998}, they show how the market dynamics, for example as measured by the Hurst exponent, can
differ with changes in traders' heterogeneous behavior.  As reviewed in Section \ref{sec_interactions}, physicists have also made contributions to
agent-based financial markets.  The order-driven agent-based financial market used by Yi-ping Huang, Shu-Heng Chen, Min-Chin Hung and Tina Yu
is, in effect, initiated by the physicist Doyne Farmer.  Their paper ``Liquidity Cost of Market Orders in the Taiwan Stock Market: A Study Based on
an Order-Driven Agent-Based Artificial Stock Market'' uses high-frequency trading data from the Taiwan Stock Exchange to simulate
the liquidity cost of market orders, which provides an alternative approach for dealing with algorithmic trading.  The paper ``Effects of Taxation
on Money Distribution'' by Marcio Diniz and Fabio Macedo Mendes extends a kind of the kinetic model of wealth distribution by taking into account
the possible influence of taxation.

\section{Concluding Remarks}

       At the end of this editorial guide, we would like to go back to the question with which
we began: what is econophysics and who are econophysicists?  From what has been presented
here, a few remarks easily stand out. First, econophysics is not limited to physicists only.
The definition of econophysics is better regarded as an intellectual one rather than a
sociological one.\footnote{See Barkley Rosser's remark \citep{rosser_2006} on the
definition of econophysics given in \citep{mantegna_stanley_2000}.}

Second, econophysics does
not just concern the application of statistical physics.  It may not be limited to physics at all.  While
statistical physics is very much the dominant force in the current development of this field, both from a
historical viewpoint and an evolutionary sociological viewpoint, this delineation is too restrictive.
It thus remains an interesting topic for economists and physicists as they review how classical
physics has shaped the later development of neoclassical economics and some of its remaining influences.
In addition, while modern econophysics is very much motivated by the recent progress in
statistical physics on scaling, universality and renormalization, and many econophysics
models, such as the Ising model, have a physical intellectual origin, it is still important to keep
the door open to contain intellectual origins from other disciplines. The minority game or the
Keynes's beauty contest, for example, obviously has an economic origin, and social networks
initiated by sociologists, together with the advancement of
modern and applied mathematics were much more independently developed before becoming the language of physics.

Third, econophysics is not just about finance. It is true that modern econophysics
is very much finance-oriented.  The first few books or textbooks on econophysics all have
``finance'', ``financial markets'', or ``speculation'' as part of their titles
\citep{osborne_1977,mantegna_stanley_2000,roehner_2002,mccauley_2004,challet_2005,voit_2005}, but there
are many other books that do not have finance as part of their titles or as their only concerns
\citep{aoyama_2010,helbing_2010}.  What is particularly evident is that many models of interactions, as we reviewed in
Section \ref{sec_interactions}, do rest upon behavioral assumptions involving other disciplines in the social sciences, such as
anthropology, sociology, psychology and game theory.  In addition, as reviewed in Section \ref{sec_networks}, econophysics has been
extensively extended to macroeconomics, international
economics, industrial economics and managerial economics. The social network analysis applied to various economic and social networks
should have good potential to be applied to interpersonal relationships in organizations.  The statistical mechanics of
networks may shed light on the psychology of networks and enhance our understanding of
the powers, reputations and the leadership of individuals in organizations \citep{kilduff_2008}.
It is then interesting to see how
econophysics may constantly expand over time from just financial markets to other branches of
economics, in particular, international macroeconomics, if the recent
financial crisis becomes one of the main concerns of econophysicists \citep{stanley_2008}.  In
this sense, a bridge will be built across the turbulent current.

\section*{Acknowledgements}

       We are grateful to Jonathan Batten, the editor of the {\em International Review of Financial
Analysis}, for his kind support in publishing a special issue for the conference {\em Econophysics
Colloquium 2010}.  We are also grateful to the National Science Council (grant no. NSC99-2916-I-001-002-A1) and
APCTP (Asia Pacific Center for Theoretical Physics) for sponsoring this conference.  The quality of this
special issue could not have been ensured without the painstaking efforts of a number of referees to whom our
heartfelt thanks are also extended. They are Chi-Ning Chen, Chung-I Chou, Yih-chyi Chuang, Herbert Dawid,
Umberto Gostoli, Serge Hayward, Tony He, Woo-Sung Jung, Mak Kaboudan, Honggang Li, John McCombie,
You-Hsien Shiau, Giulia Rotundo, Chueh-Yung Tsao, Sun-Chong Wang, Duo Wang, You-Gui Wang, Chun-Chou Wu,
Ming-Chya Wu, Ryuichi Yamamoto, and Wei-Xing Zhou.

\bibliographystyle{model1b-num-names}

%% not want to use model1b-num-names.bst.


\end{document}